\documentclass[10pt,journal,compsoc]{IEEEtran}
\usepackage{mdframed}
\usepackage{makecell}
\usepackage{csquotes}   
\usepackage{enumerate}
\usepackage{caption}
\usepackage{amssymb}
\usepackage[hyphens]{url}

\usepackage{enotez}
\usepackage{hyperref}
\let\footnote=\endnote

\DeclareInstance{enotez-list}{custom}{paragraph}
{
	heading=,
}

\usepackage{subcaption}
\usepackage{fancybox}
\usepackage{booktabs, caption, tabularx, makecell}

\usepackage[depth=1]{bookmark}
\mdfsetup{frametitlealignment=\center}

\newcommand{\hypobox}[1]{\begin{center}%
		\noindent\thicklines\setlength{\fboxsep}{8pt}%
		\cornersize{0.0}\Ovalbox{\begin{minipage}{3in}%
				#1\end{minipage}} \end{center}}

\ifCLASSOPTIONcompsoc
\usepackage[nocompress]{cite}
\else
\usepackage{cite}
\fi
\ifCLASSINFOpdf
\usepackage[pdftex]{graphicx}
\graphicspath{{./images/}}
\DeclareGraphicsExtensions{.png,.eps,.pdf}
\else
\usepackage[dvips]{graphicx}
\graphicspath{{./images/}}
\DeclareGraphicsExtensions{.png,.eps}
\fi
\AtBeginDocument{}%
\begin{document}
%
\title{An Empirical Study of Yanked Releases \\ in the Rust Package Registry}

\author{Hao Li,
	Filipe R. Cogo,
	Cor-Paul Bezemer
	\IEEEcompsocitemizethanks{\IEEEcompsocthanksitem Hao Li and Cor-Paul Bezemer are with the Analytics of Software, GAmes And Repository Data (ASGAARD) Lab, University of Alberta, Edmonton, AB, Canada. Email: li.hao@ualberta.ca, bezemer@ualberta.ca.%
		\IEEEcompsocthanksitem Filipe R. Cogo is with the Centre for Software Excellence at Huawei, Canada. Email: filipe.roseiro.cogo1@huawei.com. This work is not related to his role at Huawei.}%
	}

\IEEEtitleabstractindextext{%
	\begin{abstract}
		Cargo, the software packaging manager of Rust, provides a yank mechanism to support release-level deprecation, which can prevent packages from depending on yanked releases. Most prior studies focused on code-level (i.e., deprecated APIs) and package-level deprecation (i.e., deprecated packages). However, few studies have focused on release-level deprecation. In this study, we investigate how often and how the yank mechanism is used, the rationales behind its usage, and the adoption of yanked releases in the Cargo ecosystem. Our study shows that 9.6\% of the packages in Cargo have at least one yanked release, and the proportion of yanked releases kept increasing from 2014 to 2020. Package owners yank releases for other reasons than withdrawing a defective release, such as fixing a release that does not follow semantic versioning or indicating a package is removed or replaced. In addition, we found that 46\% of the packages directly adopted at least one yanked release and the yanked releases propagated through the dependency network, which leads to 1.4\% of the releases in the ecosystem having unresolved dependencies.
	\end{abstract}
	
	}

\maketitle
\IEEEdisplaynontitleabstractindextext
\IEEEpeerreviewmaketitle



\IEEEraisesectionheading{\section{Introduction}\label{sec:introduction}}


\IEEEPARstart{R}{ust} is a programming language that focuses on developing reliable and efficient system-level applications~\cite{rust:intro}. The official package manager of \texttt{Rust} is \texttt{Cargo}\footnote{\url{https://doc.rust-lang.org/cargo}} and the official package registry is \texttt{crates.io}\footnote{\url{https://crates.io}} which was published online in 2014. \texttt{Cargo} is considered relatively young compared to \texttt{Maven}\footnote{\url{https://maven.apache.org}} for \texttt{Java}~(2004), \texttt{Rubygems}\footnote{\url{https://rubygems.org}} for \texttt{Ruby}~(2004), \texttt{npm}\footnote{\url{https://www.npmjs.com}} for \texttt{JavaScript}~(2010), and \texttt{Packagist}\footnote{\url{https://packagist.org}} for \texttt{PHP}~(2012). 


In a software ecosystem, deprecation can happen in APIs, releases, and packages. Usually, the owner of a deprecated API plans to remove this API in the future and attempts to warn the developers who are using these APIs. For example, the owner of a package plans to remove a \texttt{foo()} function in one year, and adds a warning message which will be printed when \texttt{foo()} is called, giving a developer time to deal with the prospective deprecation. However, deprecation of releases and packages usually takes place unexpectedly. For example, when an owner of a package finds a critical bug in a release which was published a year ago, the owner can immediately deprecate this buggy release to prevent developers from using it, if the package manager supports release-level deprecation.


Prior studies have focused on the deprecation of APIs~\cite{deprecation:smalltalk} \cite{deprecation:java_language_feature} \cite{deprecatedapi:migration} \cite{deprecatedapi:python} \cite{deprecatedapi:doc_migration} \cite{deprecation:restapi} and packages~\cite{unmaintained:core_devs} \cite{unmaintained:why_fail} \cite{github:failed} \cite{unmaintained:identify} \cite{unmaintained:take_breaks} \cite{deprecation:package_MI} \cite{unmaintained:identify_failed}. In our prior work~\cite{deprecation_release:Filipe}, we studied the release-level deprecation mechanism in \texttt{npm}. \texttt{npm} and \texttt{Cargo} are the only two packaging ecosystems that have supported release-level deprecation for a long time. Other ecosystems have started to support release-level deprecation in 2019\footnote{\url{https://devblogs.microsoft.com/nuget/deprecating-packages-on-nuget-org/}}~(\texttt{NuGet}) and 2020\footnote{\url{https://discuss.python.org/t/pep-592-support-for-yanked-files-in-the-simple-repository-api/1629}}~(\texttt{PyPI}) but there is not much data available for those ecosystems yet. In this study, we focus on \texttt{Cargo}, which implements a forceful release-level deprecation~(i.e., yanking). The yank mechanism in \texttt{Cargo} will remove yanked releases from the registry index, compared to the deprecation mechanism in \texttt{npm} which just provides warning messages for deprecated releases. This forceful deprecation mechanism in \texttt{Cargo} can lead to unresolved dependencies of certain releases of a package. In addition, \texttt{Cargo} records the date on which a release was yanked, which allows us to study how the number of deprecated releases evolves~(in contrast to our prior work~\cite{deprecation_release:Filipe} in which we had to estimate the date of deprecation).

Our study mines and analyzes \texttt{Rust}'s official package registry~(\texttt{crates.io}) to improve the understanding of the yank mechanism. In addition, we compare the results in \texttt{Cargo} with \texttt{npm}. Our results can help researchers to understand release-level deprecation in software packaging ecosystems, help the \texttt{Rust} community to improve the yank mechanism, and help other packaging ecosystems to improve their own deprecation mechanism. We collected data of all the 48,823 packages on \texttt{crates.io} and used this dataset to answer the following research questions (RQs):

\begin{enumerate}[\bfseries RQ1.]
\item \textbf{How many releases are yanked?} 

The number of releases has increased steeply from 2014 to 2020 and the proportion of yanked releases also keeps increasing. In addition, we found that \texttt{Cargo} has a slightly higher proportion of yanked releases than \texttt{npm}.


\item \textbf{Why do packages use the yank mechanism?}

We study five patterns of yanking releases and we found that releases are being yanked for another reason than being defective. However, we noticed that only 5.3\% of packages explained why a release was yanked: this makes yanked releases even harder to deal with for developers who adopted such releases.

\item \textbf{How many packages adopt yanked releases?}

Even though the proportion of yanked releases is small in the ecosystem, a relatively large proportion of packages adopt yanked releases. Also, we found that yanked releases were transitively adopted in the ecosystem and caused unresolved dependencies, which in turn led to 4,158 broken releases currently in the ecosystem.

\end{enumerate}




\textbf{Paper Organization.} The rest of this paper is organized as follows. Section~\ref{sec:Background} provides background information about the package manager of \texttt{Rust} and its yank mechanism. Section~\ref{sec:related_work} discusses related work. Section~\ref{sec:methodology} presents the method that we used in our study. Section~\ref{sec:results} presents the findings of our three research questions. Section~\ref{sec:implications} discusses the implications of our findings. Section~\ref{sec:validity} discusses the threats to the validity of our study. Section~\ref{sec:conclusion} concludes this paper.

\section{Background} \label{sec:Background}

In this section, we describe how \texttt{Rust} manages packages, and we discuss the dependency requirements and yank mechanism in \texttt{Cargo}.

%

\subsection{Package management in Rust} \label{bg:Cargo}

\texttt{Cargo} is the official package manager of \texttt{Rust}. Most developers use \texttt{Cargo} to compile their packages instead of using the compiler \texttt{rustc}\footnote{\url{https://doc.rust-lang.org/rustc}} directly. Before performing a compilation, \texttt{Cargo} will resolve the dependencies and download specific versions of packages to satisfy the dependency requirements. After that, developers can run their package locally (as a standalone application) or publish their package to a package registry (as a library) using \texttt{Cargo}. 

The \texttt{Rust} community's package registry is \texttt{crates.io}, which stores the packages online and provides a platform to search and browse the information of uploaded packages. Usually, developers interact with \texttt{crates.io} through the command-line interface of \texttt{Cargo}. For example, developers can use the \texttt{cargo search}\footnote{\url{https://doc.rust-lang.org/cargo/commands/cargo-search.html}} command to find packages in \texttt{crates.io}. In addition, package owners can publish releases to \texttt{crates.io} and manage their packages through the command-line interface. For instance, \texttt{cargo publish}\footnote{\url{https://doc.rust-lang.org/cargo/commands/cargo-publish.html}} will upload the current package to a registry~(which is set to \texttt{crates.io} by default). In this paper, for simplicity we refer to \texttt{crates.io} as \texttt{Cargo}. 

%
%

\subsection{Dependencies in Cargo} \label{bg:deps}

Dependency requirements in \texttt{Cargo} are based on semantic versioning\footnote{\url{https://github.com/steveklabnik/semver}} and \texttt{Cargo} will determine the version of dependencies when developers build their projects. The semantic versioning specification defines that a version number consists of three parts: major, minor, and patch. For example, version number \texttt{1.2.3} has a major number \texttt{1}, a minor number \texttt{2}, and a patch number \texttt{3}. The packages should guarantee that patch updates only introduce \enquote{backwards compatible bug fixes}, minor updates only add features which are backwards compatible, and only the major updates can introduce breaking changes. We refer to this guarantee as the \textit{semantic versioning guarantee} in our paper.

The semantic versioning specification is used by \texttt{Cargo.toml},\footnote{\url{https://doc.rust-lang.org/cargo/reference/manifest.html}} a file under the directory of a \texttt{Rust} project, to store the dependency requirements. The interpretation of requirement statements is different across software ecosystems~\cite{ecosystems:semver}. Table~\ref{tab:dep_reqs} shows the versioning specifications which are used in \texttt{Cargo}. It is notable that \texttt{Cargo} interprets \texttt{1.2.3} as a caret requirement~(\texttt{$\wedge$1.2.3}) and the wildcard requirement statement \enquote{$\star$} (i.e., matching any version) was banned in January 2016.\footnote{\url{https://doc.rust-lang.org/cargo/faq.html}} \texttt{Cargo} searches the registry index to find versions which can satisfy the requirements and downloads dependencies. If there are multiple versions available that satisfy a requirement, \texttt{Cargo} will choose the version which has the largest version number. We call the owner of a dependency requirement a \textit{client} and the package to which the dependency requirement points a \textit{provider}. For example, a client package \texttt{C} has a dependency requirement \texttt{3.0.1} for a release from a provider package \texttt{P}. \texttt{Cargo} will choose the greatest version \texttt{3.5.1} from \texttt{P} which satisfies this requirement even though there exists an exactly matched version \texttt{3.0.1}~(since \texttt{Cargo} interprets the requirement \texttt{3.0.1} as a caret requirement).

\begin{table}[t]	
	\centering	
	\caption{Five types of versioning specifications in Cargo}
	\label{tab:dep_reqs}
	\begin{tabular}{ l l l} 
		\toprule
		\textbf{Types} & \textbf{Statement} & \textbf{Interpretation} \\ 
		\midrule
		\textbf{Comparison} 	& $=$1.2.3 		& [1.2.3]  \\ 
		                    	& $>$1.2.3 		& ]1.2.3, $+\infty$[  \\ 
		                    	& $<$1.2.3 		& [0.0.0, 1.2.3[  \\ 
		                    	& $\ge$1.2.3	& [1.2.3, $+\infty$[  \\ 
		\textbf{Compound} 		& $>$1.2.3, $\le2.3.4$ 	& ]1.2.3, 2.3.4]  \\ 
		\textbf{Caret} 			& $\wedge$1 	& [1.0.0, 2.0.0[  \\ 
		 						& $\wedge$1.2 	& [1.2.0, 2.0.0[  \\ 
		 						& $\wedge$1.2.3	& [1.2.3, 2.0.0[  \\
		 						& $\wedge$0 	& [0.0.0, 1.0.0[  \\
		 						& $\wedge$0.1 	& [0.1.0, 0.2.0[  \\
		 						& $\wedge$0.0.1 & [0.0.1, 0.0.2[  \\
		 						& $\wedge$0.1.2 & [0.1.2, 0.2.0[  \\
		\textbf{Tilde} 			& $\sim$1 	        & [1.0.0, 2.0.0[  \\ 
								& $\sim$1.2 	        & [1.2.0, 1.3.0[  \\ 
								& $\sim$1.2.3		& [1.2.3, 1.3.0[  \\ 
		\textbf{Wildcard} 		& $\star$\footnotesize{$^a$} 			& [0.0.0, $+\infty$[  \\ 
								& 1.$\star$ 			& [1.0.0, 2.0.0[  \\ 
								& 1.2.$\star$ 		& [1.2.0, 1.3.0[  \\ 
								& 1.2.3 		& [1.2.3, 2.0.0[  \\ 
		\bottomrule
		\footnotesize{$^a$: removed in 2016} 
	\end{tabular}
\end{table}

%


\subsection{Yanked releases} \label{bg:yanked}

\texttt{Cargo} provides a command called \texttt{cargo yank}\footnote{\url{https://doc.rust-lang.org/cargo/commands/cargo-yank.html}} to deprecate a published release, which can also be unyanked with the \texttt{yank undo} command. After a developer calls the yank command for a certain release, this release will be indicated as yanked and is no longer available from the registry index for \texttt{Cargo}. Thus, when \texttt{Cargo} is trying to resolve dependencies for a project, it will automatically skip yanked releases and choose the release which has the largest version number that still satisfies the dependency requirement. For example, a client package \texttt{C} has a dependency requirement \texttt{$\sim$2.5.1} for a package \texttt{P}. The latest releases of \texttt{P} are \texttt{2.5.5} and \texttt{2.5.6}. However, the latter was yanked. Hence, \texttt{Cargo} will select release \texttt{2.5.5} of \texttt{P}. Due to the deletion of yanked releases from the registry index, \texttt{Cargo} cannot download a yanked release even if the dependency requirement uses the \enquote{=} operator.


Notably, the yank command does not completely delete any data from the package registry, hence the yanked releases can still be downloaded through work arounds. One approach is using the download API which is provided by the package registry,\footnote{\url{https://doc.rust-lang.org/cargo/reference/registries.html}} and another approach is through the locking mechanism of \texttt{Cargo}. \texttt{Cargo} will generate a \texttt{Cargo.lock}\footnote{\url{https://doc.rust-lang.org/cargo/guide/cargo-toml-vs-cargo-lock.html}} file if the building process is successful. This file stores the versions of dependencies that were used during the build. When the developer compiles the project a second time, \texttt{Cargo} will reuse the versions of the dependencies that are stored in \texttt{Cargo.lock} (as long as the developers did not change a required version in \texttt{Cargo.toml}) even if the depended versions are yanked or a newer version is available. A standalone application will usually upload both \texttt{Cargo.toml} and \texttt{Cargo.lock} to its repository,\footnote{\url{https://doc.rust-lang.org/cargo/faq.html#why-do-binaries-have-cargolock-in-version-control-but-not-libraries}} which assures the reproducibility of the building process. In contrast, a library will upload \texttt{Cargo.toml} but not \texttt{Cargo.lock}, so \texttt{Cargo} will help the clients of this library to determine a suitable version to use. However, as we show in Section~\ref{res:yanked_adoption}, this could lead to unresolved dependencies when building a package that depends on a yanked release.

\section{Related work} \label{sec:related_work}

In this section, we discuss related work about software packaging ecosystems and the deprecation of APIs and packages.

\subsection{Software packaging ecosystems}

Most research on software packaging ecosystems has focused on \texttt{npm}~\cite{npm:downgrades}\cite{npm:security_vulnerabilities}\cite{npm:ecosys}\cite{npm:popularity_metrics} of \texttt{JavaScript}, \texttt{PyPI}~\cite{pypi:search}\cite{pypi:dormancy}\cite{pypi:watchman} of \texttt{Python}, and \texttt{CRAN}~\cite{deprecation:R_dashboard}\cite{deprecation:R_maintain}\cite{ecosystem:github_cran}\cite{deprecation:R_evolution} of \texttt{R}. In this paper, we study the yank mechanism in the \texttt{Cargo} ecosystem, the packaging system of \texttt{Rust}.

Few studies have focused on the \texttt{Cargo} ecosystem. Evans et~al.~\cite{rust:unsafe} studied the safety of packages in the \texttt{Cargo} packaging ecosystem of \texttt{Rust}. They found that 29\% of the packages directly use the \texttt{unsafe} keyword, which is provided by \texttt{Rust} to avoid safety checking of the compiler. Furthermore, they observed that popular packages use \texttt{unsafe} more frequently. 

Many studies include \texttt{Cargo} as a subject when comparing multiple software packaging ecosystems. Decan and Mens~\cite{ecosystems:zeroversion} investigated pre-releases of three packaging ecosystems and observed that more than 90\% of the packages in \texttt{Cargo} published a pre-release as their latest release. In addition, they found that most dependencies that point to pre-releases allow patch updates in \texttt{Cargo}~\cite{ecosystems:semver}, which does not follow the semantic versioning specification. Constantinou et~al.~\cite{ecosys:cross_sys_packages} studied packages which are distributed across multiple packaging ecosystems and found that these packages in \texttt{Cargo} have more stars on \texttt{GitHub} than in other ecosystems. Like other packaging systems, a relatively small proportion of packages are depended on by most of the packages in \texttt{Cargo}~\cite{dependency:Decan}.




Many researchers have studied the \texttt{npm} ecosystem. In prior work~\cite{npm:downgrades}, we investigated dependency downgrades in \texttt{npm} and observed that packages changed their dependency constraints for migrating away from defective dependencies. Decan et~al.~\cite{npm:security_vulnerabilities} also found that a proper dependency constraint can help a package migrate away quickly from a vulnerable dependency. In addition, they found that most of the security vulnerabilities are fixed before they are published in \texttt{npm}. Wittern et~al.~\cite{npm:ecosys} studied dependencies in \texttt{npm} and found that the package dependencies keep increasing. However, most of the dependencies point to a small proportion of packages in the ecosystem. Zerouali et~al.~\cite{npm:popularity_metrics} analyzed various popularity metrics in \texttt{npm} and found that the results of identifying popular packages can be different based on the metrics used.


Imminni et~al.~\cite{pypi:search} implemented a semantic search engine for the \texttt{PyPI} ecosystem since \texttt{PyPI} has a limited ability to provide quality search results for developers. To detect dependency conflicts in \texttt{PyPI}, Wang et~al.~\cite{pypi:watchman} developed a tool to monitor the ecosystem. Valiev et~al.~\cite{pypi:dormancy} built survival models for \texttt{PyPI} to analyze the risk of a package become dormant.

German et~al.~\cite{deprecation:R_evolution} studied the \texttt{CRAN} packaging ecosystem of \texttt{R}. They found that most dependencies point to popular packages and user-contributed packages need more time to grow their community than core packages. Claes et~al.\cite{deprecation:R_maintain} observed that the time of fixing errors in \texttt{CRAN} packages differs across operating systems. They also~\cite{deprecation:R_dashboard} developed a tool to analyze the maintainability of a package in \texttt{CRAN}, which can visualize information such as release history, dependencies and namespace. Decan et~al.~\cite{ecosystem:github_cran} found that packages in \texttt{CRAN} also manage their repositories on \texttt{Github}, which can influence the dependency management.




\begin{figure*}[t]
	\centering
	\includegraphics[width=6.5in]{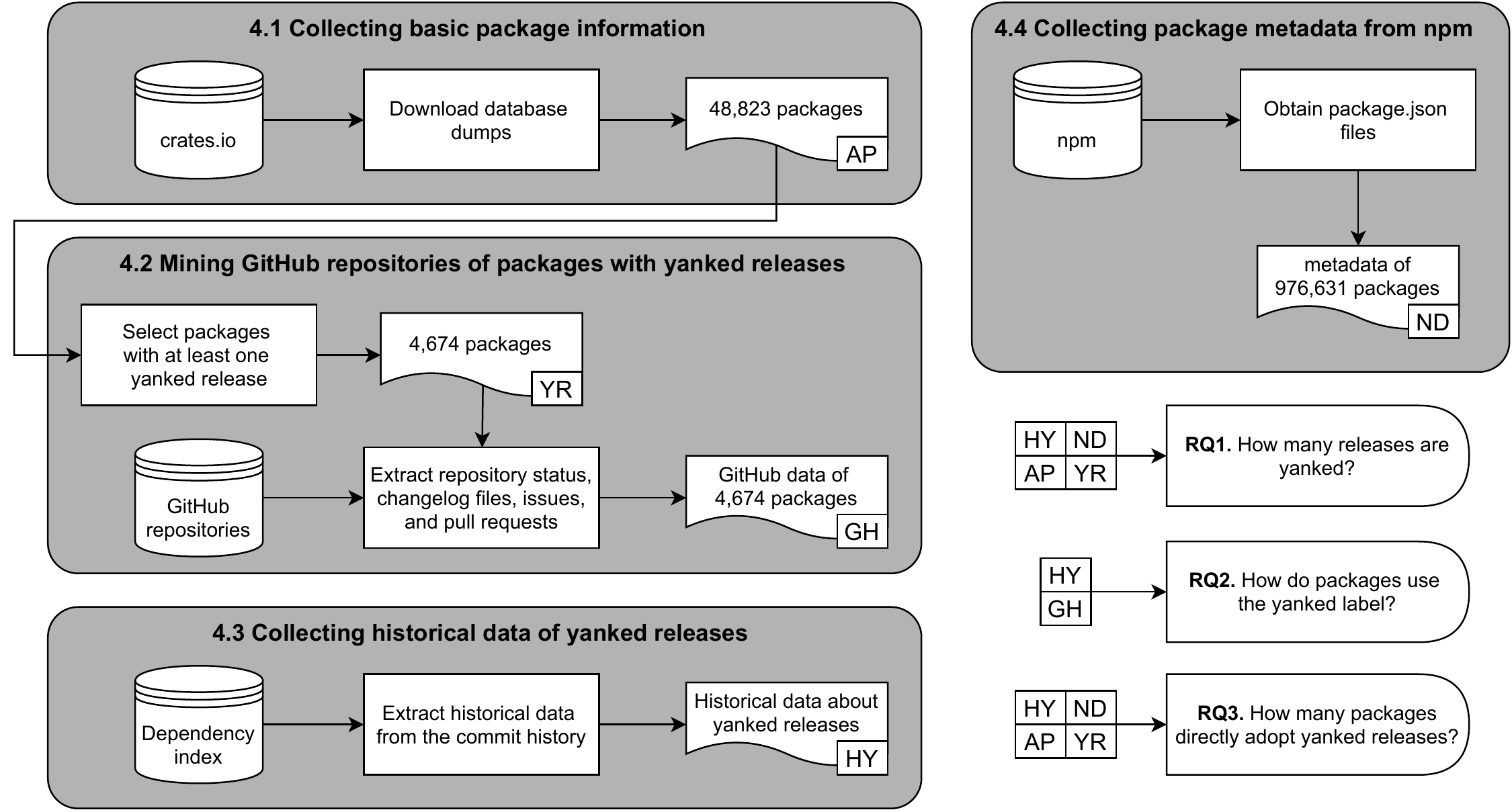}
	\caption{Overview of our methodology.}
	\label{fig:methodology}
\end{figure*}
\subsection{Deprecated APIs and packages}


In prior work~\cite{deprecation_release:Filipe}, we studied the deprecation mechanism in \texttt{npm}. To the best of our knowledge, that was the first study that focused on the release-level deprecation mechanism of a software packaging ecosystem. This follow-up paper focuses on the yank mechanism in \texttt{Cargo} and compares it with the deprecation mechanism in \texttt{npm}. The reason is threefold: 1) The yank mechanism in \texttt{Cargo} is more forceful than the deprecation mechanism in \texttt{npm}. 2) \texttt{Cargo} provides the date of yanking which supports a more in-depth analysis. 3) \textit{\enquote{Comparative studies can be seen as a prerequisite for designing successful domain-specific ecosystem solutions}}~\cite{ecosys:challenges}. Hence, the comparisons across these two software ecosystems can help us better understand the design of a release-level deprecation mechanism.


Many researchers have studied deprecated APIs at the code-level. Sawant et~al.~\cite{deprecation:java_language_feature} interviewed \texttt{Java} API producers and surveyed \texttt{Java} developers, and suggested \texttt{Java} to provide a warning mechanism for developers. Wang et~al.~\cite{deprecatedapi:python} investigated six popular packages in \texttt{Python} and observed that developers did not have a consistent strategy to deprecate an API, and that about 25\% of the deprecated APIs are not documented. Robbes et~al.~\cite{deprecation:smalltalk} analyzed deprecated functions and classes in \texttt{Smalltalk}, and found that about half of the deprecation messages cannot help the developers to migrate away from the deprecation. 


Few studies have focused on the deprecation of web APIs. Yasmin et~al.~\cite{deprecation:restapi} analyzed 1,368 RESTful APIs and found that most of the removed APIs did not deprecate the interface to inform their users before introducing the deletion.



Unlike deprecated APIs, it is not easy to identify whether a package is deprecated because the owner may not indicate deprecation in the documentation. Coelho et~al.~\cite{unmaintained:identify} found that the important features to predict whether a package is deprecated or unmaintained include the number of commits and closed issue reports. Khondhu et~al.~\cite{deprecation:package_MI} introduced the maintainability index to identify whether a package is inactive or abandoned on \texttt{SourceForge.net}. In contrast, Maqsood et~al.~\cite{unmaintained:identify_failed} implemented eight machine learning algorithms to identify successful projects. 

Many researchers conducted studies to understand the reasons behind the deprecated and abandoned packages. Coelho et~al.~\cite{github:failed} surveyed the owners of 104 deprecated \texttt{GitHub} packages, and showed that the reasons include environmental factors, project characteristics, and human factors. Iaffaldano et~al.~\cite{unmaintained:take_breaks} interviewed developers from the open-source software community, and also found the reasons behind abandoned packages include human factors and project characteristics. Avelino et~al.~\cite{unmaintained:core_devs} found that the loss of core developers can increase the risk of a package becoming abandoned.

\section{Methodology} \label{sec:methodology}

In this section, we introduce the methodology of our study of yanked releases in the \texttt{Rust} package registry. Figure~\ref{fig:methodology} gives an overview of our study.


\subsection{Collecting basic package information}

\begin{table}[t]
	\caption{Key information in the database.}
	\label{tab:database_table}
	\begin{tabularx}{\linewidth}{lX}
		\toprule
		\textbf{Field} & \textbf{Description} \\ 
		\midrule
		versions.id & The identifier of a release.  \\
		versions.num & The semantic version number of a release such as \texttt{1.2.3} or \texttt{0.1.2-alpha}. \\
		versions.created\_at & The creation date of a release.  \\
		versions.yanked & A flag to indicate whether a release is yanked.  \\
		crates.id & The identifier of a package. \\
		crates.readme & The content of the \texttt{readme} file in a package.  \\
		crates.repository & The link to the repository of a package.  \\
		dependencies.version\_id & The identifier of the release to which the dependency belongs.  \\
		dependencies.crate\_id & The identifier of the package to which the dependency points.  \\
		dependencies.req & The dependency requirement (e.g., \texttt{$\wedge$1.2.3} or \texttt{$\sim$1.2.3}).  \\
		\bottomrule
	\end{tabularx}
\end{table}


\texttt{Cargo} provides database dumps\footnote{\url{https://crates.io/data-access}} which contain all the information (e.g., dependencies, downloads, creation date) exposed through the official API. The database dumps are the primary data source of our study and we downloaded the dump that contains the information of 48,823 packages with 294,801 releases on October 29th, 2020. Table~\ref{tab:database_table} shows the database fields which store important information for our study.

\subsection{Mining GitHub repositories of packages with yanked releases}

As Table~\ref{tab:database_table} shows, yanked flags are stored in the \texttt{versions} table. We retrieved all entries that are indicated as yanked. Then, we selected the packages which have at least one yanked release and collected the links to their repositories. For the links which direct to a \texttt{GitHub} repository, we used the \texttt{GitHub} API\footnote{\url{https://docs.github.com/en/rest}} to extract the issue reports and pull requests of these repositories. In addition, we collected the status of these repositories (active, archived or forked) through the \texttt{GitHub} API. If the repository cannot be found, we marked its link as invalid.

Furthermore, we collected the changelogs of the packages which have at least one yanked release from their \texttt{readme} file and \texttt{GitHub} repository. The \texttt{readme} field of the \texttt{crates} table in the database contains the content of the \texttt{readme} file. We identified whether the \texttt{readme} contains a changelog by searching for the keywords \enquote{changelog}, \enquote{change log}, \enquote{release notes}, and \enquote{release note} in the content, as well as searching \enquote{news} and \enquote{history} in the headings. For packages which provide a valid link to their \texttt{GitHub} repository, we queried the filenames in the root directory of the repository and collected the file if the filename matches the same keywords which we used above.

\subsection{Collecting historical data of yanked releases}

\texttt{Cargo} determines the dependencies based on the registry index which is managed in a \texttt{Git}\cite{git} repository. This index repository\footnote{\url{https://github.com/rust-lang/crates.io-index}} contains the information~(e.g., dependencies, version numbers, and yanking flags) of all published releases. The data of each package is stored in separate files, and the information is updated automatically whenever a change occurs (e.g., a new release is uploaded, yanked or unyanked). Because all the changes are managed in the \texttt{Git} repository, the commit history contains the date of each change. We mined the commit messages to extract when a release was yanked or unyanked. 

However, we noticed that the commit history in the main branch is not complete because the maintainer of \texttt{Cargo} regularly squashed commits into one to speed up the cloning of the repository.\footnote{\url{https://internals.rust-lang.org/t/cargos-crate-index-upcoming-squash-into-one-commit}} These squashed commits are stored in \texttt{snapshot} branches, hence we collected all the commits of these branches to obtain a complete historical overview of (un)yanked releases.


%
%

\subsection{Collecting package metadata from npm}

We reused the dataset from our prior study~\cite{deprecation_release:Filipe} which contains the metadata of 976,613 packages from \texttt{npm} at May 5th, 2019. For each package, we collected the information of all releases and selected the dependencies in their latest releases. Finally, we extracted 7,829,362 releases and 6,178,019 dependencies from 976,613 packages.


\section{Results} \label{sec:results}

In this section, we present the motivation, approach, and findings for each of our three research questions~(RQs).

\subsection{RQ1: How many releases are yanked?} \label{res:how_many_yanked}


\textbf{Motivation.} Deprecation can happen at the code-level, release-level, and package-level. In our prior work~\cite{deprecation_release:Filipe}, we studied the release-level deprecation mechanism in \texttt{npm}. Similarly, \texttt{Cargo} has a yank mechanism for release-level deprecation to allow the owner of a package to \enquote{remove a previously published crate's version from the server's index}.\footnote{\url{https://doc.rust-lang.org/cargo/commands/cargo-yank.htm}l} In contrast to \texttt{npm}, \texttt{Cargo} records the date on which a release was yanked or unyanked. Hence, we can study how often developers use the yank mechanism in the history of \texttt{Cargo}. The goal of this research question is to understand the yank mechanism in \texttt{Cargo} by investigating the frequency of yanked releases and comparing the result with \texttt{npm}. 


%

\medskip\noindent\textbf{Approach.} We calculated the proportion of yanked releases and packages that have at least one yanked release in \texttt{Cargo} to measure how often the yank mechanism is used. To analyze the trend of usage, we investigated the historical information of releases and yanked releases. We collected the date on which a release was published from the \texttt{created\_at} field of the \texttt{versions} table~(as shown in Table~\ref{tab:database_table}). In addition, to count the yanked releases in a certain period more precisely, we also considered the date of unyanking a release. Finally, we calculated the number of releases and the proportion of yanked releases from November 2014 to October 2020.

Next, we calculated the yanking rate (i.e., the percentage of yanked releases in a package) for every package. For example, the yanking rate is 100\% for \textit{fully yanked} packages (i.e., packages of which all releases are yanked), and 0\% for packages which do not have any yanked release. Then, we compared our findings between \texttt{Cargo} and \texttt{npm} by performing the Mann-Whitney U test~\cite{Mann1947OnAT} at a significance level of $\alpha=0.05$ to determine whether the differences are significant. However, the Mann-Whitney U test only determines whether two distributions are different. Therefore, we computed Cliff's delta~$d$~\cite{Cliff} effect size to quantify the difference. To explain the value of~$d$, we used the thresholds which are provided by Romano et~al.~\cite{Cliff_threshold}:

$$
\mathrm{Effect \ size} = 
\left\{
\begin{array}{ll}
	negligible,  & \mathrm{if} \ |d|  \le 0.147 \\
	small,  & \mathrm{if} \ 0.147 < |d|  \le 0.33 \\
	medium,  & \mathrm{if} \ 0.33 < |d|  \le 0.474 \\
	large,  & \mathrm{if} \ 0.474 < |d|  \le 1 \\
\end{array}\right.
$$

\begin{figure}[t]
	\centering
	\includegraphics[width=3.3in]{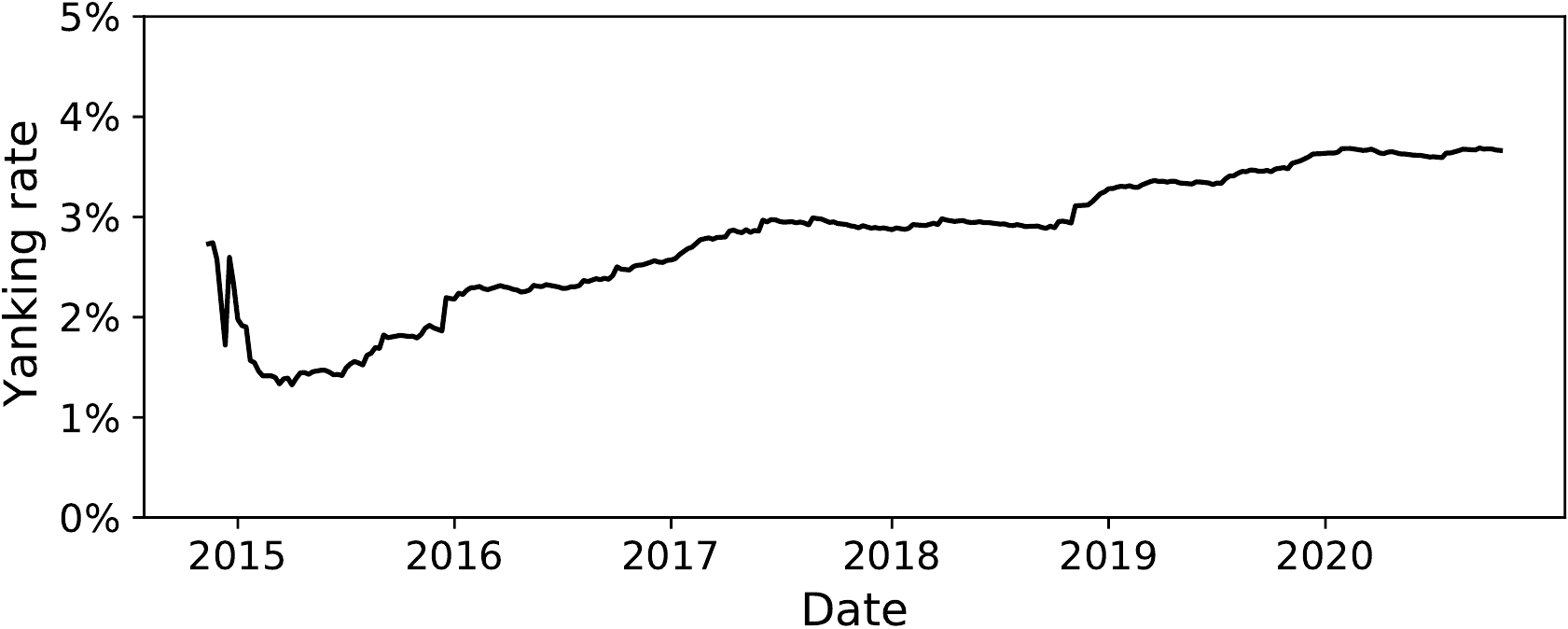}
	\caption{The percentage of yanked releases in \texttt{Cargo} from November 2014 to October 2020.}
	\label{fig:yanked_history}
\end{figure}

\medskip\noindent\textbf{Findings.} \textbf{3.7\% of the releases in Cargo are yanked and 9.6\% of the packages have at least one yanked release.} There are 10,761 yanked releases in \texttt{Cargo} and 4,674 packages with at least one yanked release. We found that the proportion of yanked releases in \texttt{Cargo} (3.7\%) is close to the proportion of deprecated releases (3.2\%) in \texttt{npm}. In contrast, \texttt{Cargo} has a larger proportion of packages which have at least one yanked release (9.6\%), compared to the proportion of packages with at least one deprecated release in \texttt{npm} (3.7\%). 

\textbf{Between 2014 and 2020, the percentage of yanked releases in Cargo has gradually increased from 1.4\% to 3.7\%.}  We found that unyanking only happened 725 times in the history, which is relatively uncommon compared to 10,761 yanked releases in \texttt{Cargo}. Figure~\ref{fig:yanked_history} shows that the cumulative number of releases in \texttt{Cargo} has increased steeply from 0 to nearly 300,000 in the period 2014 to 2020. The Cox-Stuart test~\cite{cox-stuart-test} shows that the increasing trend of the number of releases is significant~($p\ll0.05$). Also, the yanking rate has increased gradually to 3.7\% since 2014, and again the Cox-Stuart test shows that the increasing trend is significant~($p\ll0.05$). We cannot analyze the trend of the yanking rate in \texttt{npm} because it does not provide historical information about deprecated releases.




\begin{figure}[t]
	\begin{subfigure}[b]{0.3\textwidth}
		\includegraphics[width=3.3in]{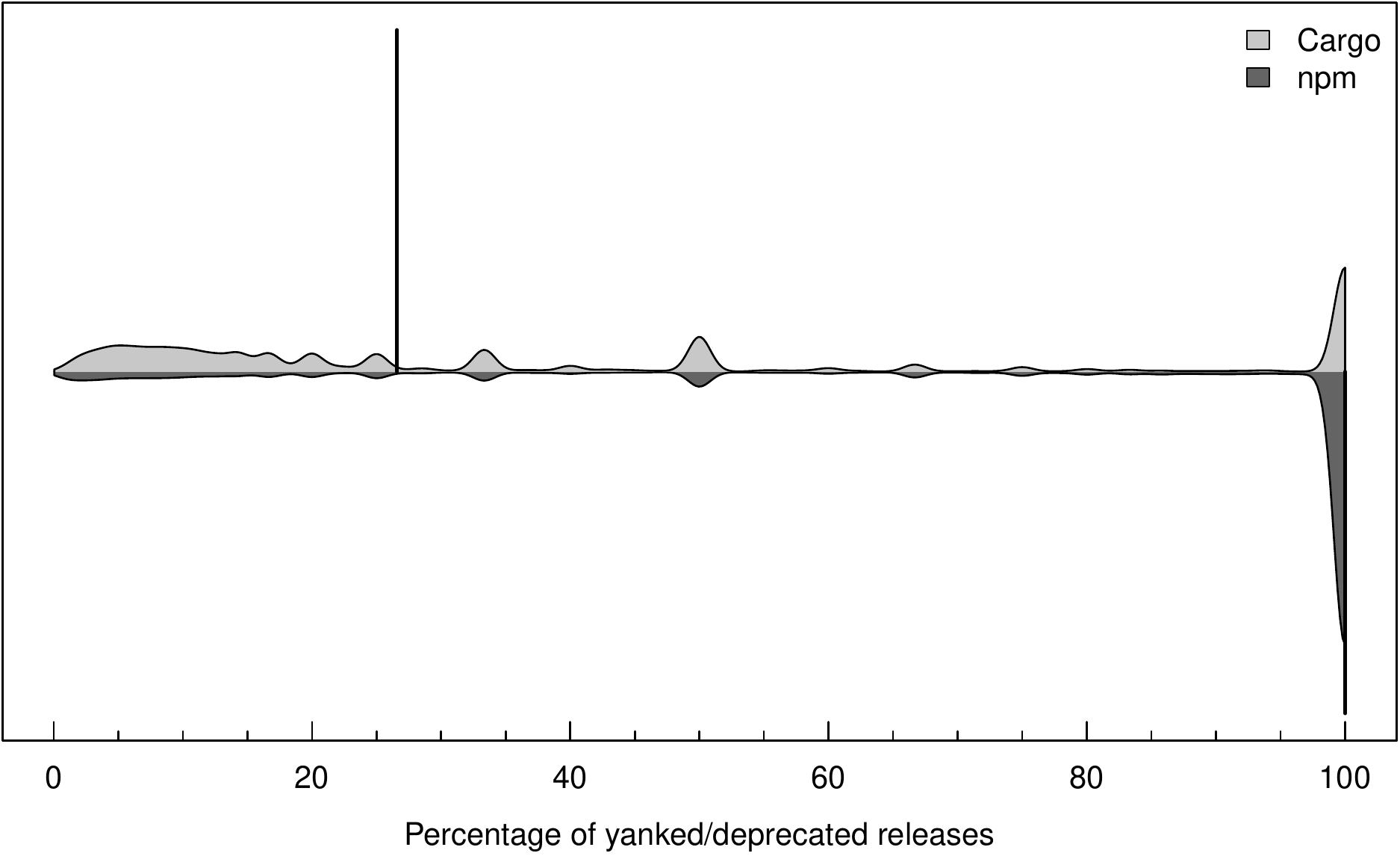}
	\end{subfigure}
	\par\smallskip
	\par\smallskip
	\par\smallskip
	\begin{subfigure}[b]{0.3\textwidth}
		\includegraphics[width=3.3in]{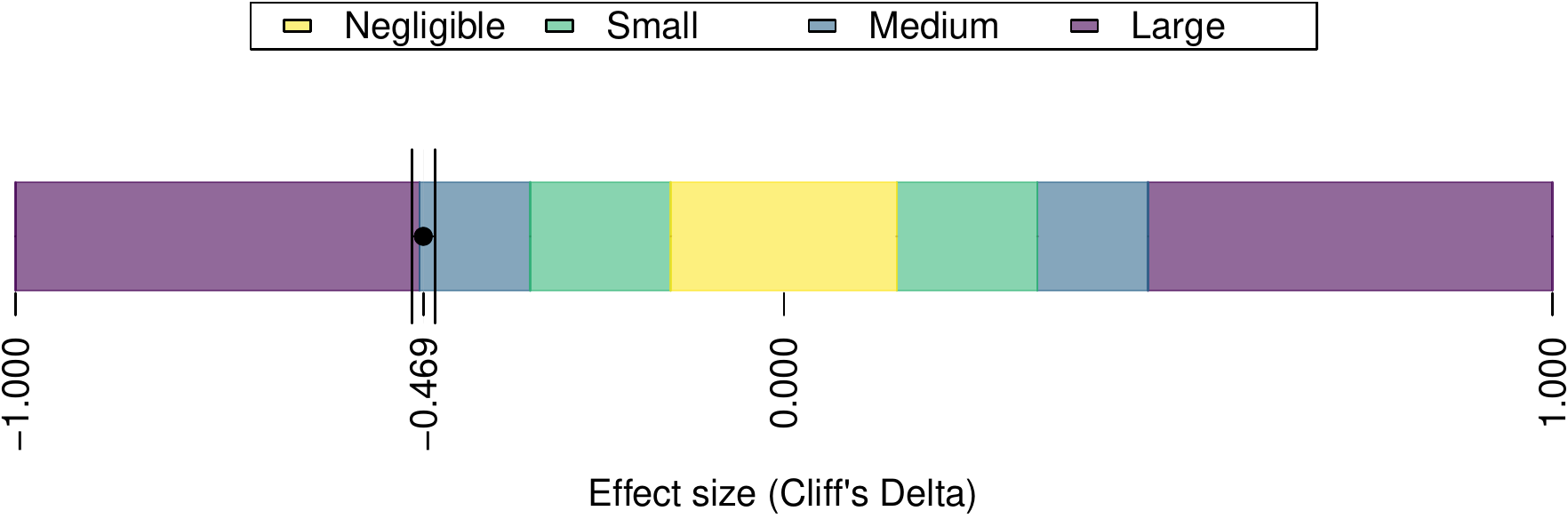}
	\end{subfigure}
	\caption{Distributions of the yanking rate of packages with at least one yanked release in \texttt{Cargo} and \texttt{npm} and Cliff's Delta~$d$.}
	\label{fig:percentage_compare}
\end{figure}

\textbf{It is more common to partially yank a package in Cargo than in npm.} The majority (75\%) of the packages with at least one deprecated release are partially yanked packages in \texttt{Cargo}. In contrast, partially deprecated packages are the minority (34\%) of packages with at least one yanked release in \texttt{npm}. Among the partially yanked packages, the median yanking rate is 17\% in \texttt{Cargo}, which is half of the median number in \texttt{npm} (33\%). Figure~\ref{fig:percentage_compare}~(using the package from Lin et~al.~\cite{dayilin:draw}) shows the distributions of the yanking rate in the two ecosystems and the value of Cliff's Delta $d$. The Mann-Whitney U test shows that the distributions of the yanking rate are significantly different in these two software ecosystems. In addition, the value of Cliff's Delta $|d|$ is 0.469, which indicates that the effect size is medium.

\hypobox{
	\textbf{RQ1 Summary:} 
	The proportions of yanked releases in \texttt{Cargo} and \texttt{npm} are similar~(3.7\% vs. 3.2\%), but it is much more common in \texttt{Cargo}~(75\% vs. 34\%) to yank only a few releases of a package.
}


\subsection{RQ2: Why do packages use the yank mechanism?} \label{res:yanked_in_practice}

\begin{figure}[t]
	\centering
	\includegraphics[width=2.8in]{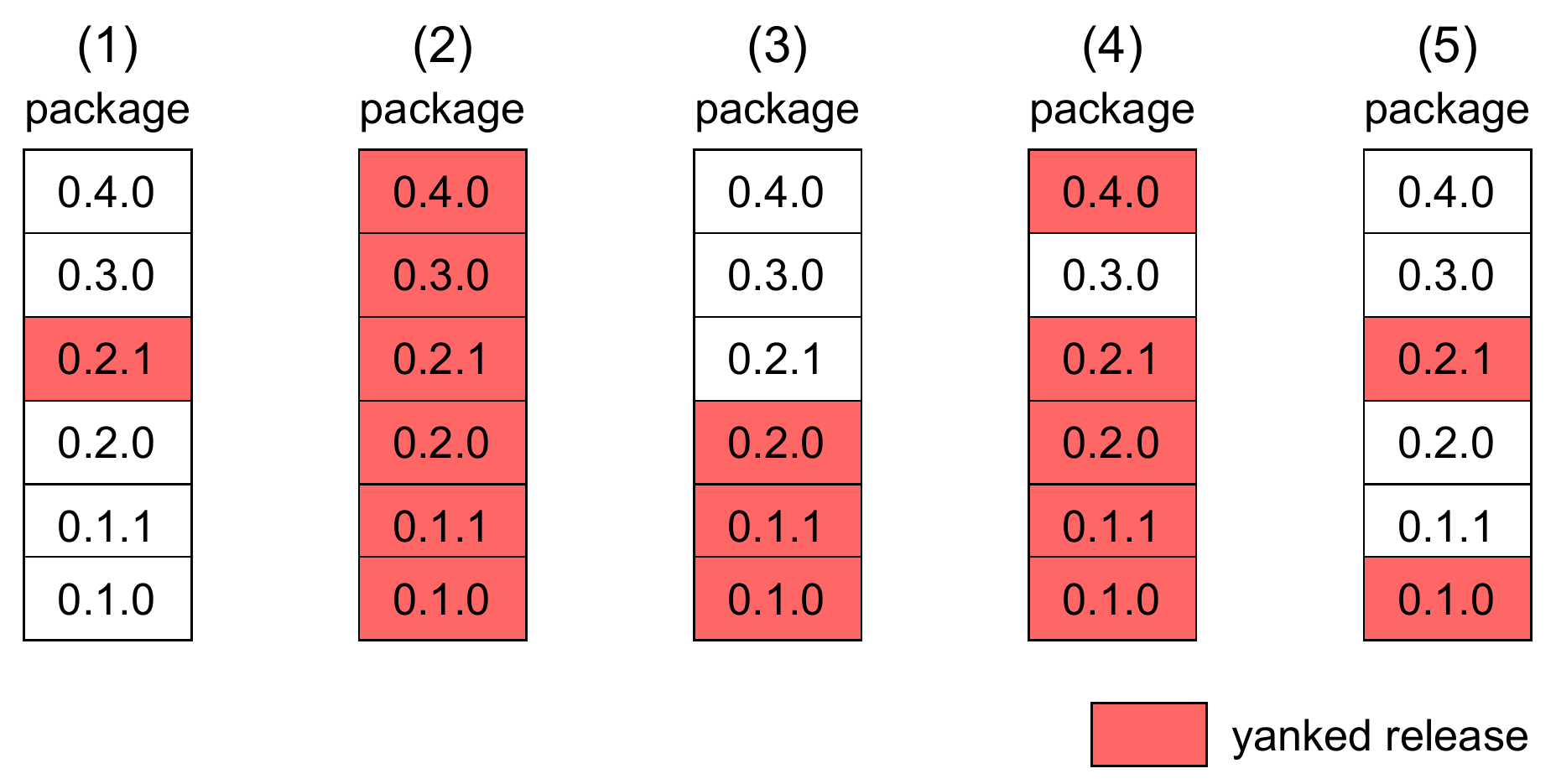}
	\caption{Five patterns of yanking: (1)~A package yanked only one release; (2)~A package yanked all releases; (3)~A package yanked back-to-back releases; (4)~A package yanked all releases except one; (5)~A package yanked nonadjacent releases.}
	\label{fig:RQ2_patterns}
\end{figure}

\textbf{Motivation.} In our prior work~\cite{deprecation_release:Filipe}, we studied non-forceful release-level deprecation in \texttt{npm}. However, in \texttt{Cargo}, release-level deprecation~(yanking) is forceful, which means that releases are no longer accessible once they are deprecated. In this research question, we take a closer look at packages with at least one yanked release in \texttt{Cargo} to investigate why the yanking mechanism is used and to study the rationales for yanking a release. The results can help us to understand the release-level deprecation in \texttt{Cargo} from the developers' point of view.
\medskip\noindent\textbf{Approach.} To understand the usage of the yank mechanism, we looked for five possible patterns in which releases are yanked in \texttt{Cargo}~(as shown in Figure~\ref{fig:RQ2_patterns}). First, we collected the packages which have at least one yanked release and sorted their releases based on the release date. Then, we went through the collected 4,674 packages to identify whether a package belongs to one of the five patterns. 

In addition, we investigated the changelogs, issue reports, and pull requests from the 4,674 packages to analyze the rationales behind yanking. We selected the packages which contain the terms \enquote{yank} or \enquote{deprecate} in their changelogs, issue reports, or pull requests. The first author went through the selected 638 packages and filtered out 380 packages which did not provide information that is related to the yanked release. After that, the first and third author performed open card sorting together to identify the rationales behind the yanked releases of the remaining 258 packages. We could not identify the rationales for 9 out of 258 packages during the card sorting~(i.e., 3.5\% false positives of the filtering process). Hence, our results cover the remaining 249 packages.

\medskip\noindent\textbf{Findings.} \textbf{The usage of the yank mechanism among packages in Cargo follows one of the five patterns in Figure~\ref{fig:RQ2_patterns}.} In addition, developers yanked releases for 11 reasons~(see Table~\ref{tab:yanking_rationales}). We summarize below the patterns that we were able to identify, together with some examples and the rationales we categorized in the card sort. 


\begin{table*}[t]	
	\caption{Identified rationales behind yanked releases in the card sort}
	\label{tab:yanking_rationales}
	\begin{tabularx}{\linewidth}{ l X r r r r r r}
		\toprule
		\textbf{Rationale}              & \textbf{Description} & \textbf{Pkg} & \textbf{P1} & \textbf{P2} & \textbf{P3} & \textbf{P4} & \textbf{P5} \\ \midrule
		Breaking SemVer                 &  The release introduces breaking changes and therefore does not follow the semantic versioning specification specification                    & 43.0\%         & 47.5\%        & 25.0\%        & 40.0\%        & -             & 42.9\%        \\
		Defect                          &  Release contains a defect/bug                    & 36.9\%         & 28.0\%        & -             & 46.7\%        & 100.0\%       & 50.0\%        \\
		Fixing \enquote{*} dependencies &  Release uses the wildcard dependency~(\enquote{*}) which has been prohibited since 2016\footnote{\url{https://doc.rust-lang.org/cargo/faq.html}}                   & 7.2\%          & 4.2\%         & 25.0\%        & 8.3\%         & -             & 8.9\%         \\
		Package removed or replaced     & The whole package is removed or is replaced by another package                     & 5.2\%          & 5.1\%         & 50.0\%        & -             & 33.3\%        & -             \\
		Broken dependencies             & The release contains a broken dependency                     & 4.8\%          & 6.8\%         & -             & 1.7\%         & -             & 5.4\%         \\
		Bump propagation                & A patch release that includes a minor/major update of an existing dependency and therefore should be a minor/major release as well                      & 3.2\%          & 5.1\%         & -             & 1.7\%         & -             & 1.8\%         \\
		MSRV\footnote{\url{https://github.com/rust-lang/rfcs/pull/2495}} policy                     & Upgrading the minimum supported rust version~(which is a breaking change) in a patch update                     & 2.8\%          & 0.8\%         & -             & 6.7\%         & -             & 3.6\%         \\
		Yanked dependencies             & Dependencies are yanked                     & 2.0\%          & 3.4\%         & -             & 1.7\%         & -             & -             \\
		Placeholder release             & An initial release for holding the name in \texttt{Cargo}                    & 1.2\%          & 0.8\%         & -             & 1.7\%         & -             & 1.8\%         \\
		License updated                 & Forgetting to update the license                     & 0.4\%          & -             & -             & -             & -             & 1.8\%         \\
		Unsupported                     & Releases are no longer supported                     & 0.4\%          & -             & -             & 1.7\%         & -             & -             \\ \bottomrule 
	\end{tabularx}
	\footnotesize{Note: one package can have multiple rationales as it can have multiple yanked releases. We identified one rationale per yanked release. The \enquote{Pkg} column presents the percentage of packages that contain the corresponding rationales. The \enquote{P1} to \enquote{P5} columns present the percentage of packages that contain the corresponding rationales under the yanking patterns. } 
\end{table*}


%



\medskip\noindent\textbf{Pattern 1: Yanking only one release (40\% of the packages).} We found this pattern in 1,879 packages and \textit{Breaking SemVer} is the main rationale behind this pattern~(47.5\%) in the card sort. One example is \texttt{pyo3}\footnote{\url{https://crates.io/crates/pyo3}}, an issue report of this project mentions that \enquote{[v]ersion 0.5.1 breaks SemVer guarantees}\footnote{\url{https://github.com/PyO3/pyo3/issues/285}} because the owner of \texttt{pyo3} accidentally merged a new feature to this patch update for \texttt{0.5.0}. It introduced breaking changes and should be a minor or major update since the semantic versioning guarantee requires that patch updates only introduce backwards-compatible bug fixes. Hence, the owner yanked \texttt{0.5.1}, backported the changes, and published \texttt{0.5.2} which did not include the new feature.


In addition, we noticed that developers yanked a release for \textit{Broken dependencies}~(6.8\%) or \textit{Yanked dependencies}~(3.4\%). One example of \textit{Broken dependencies} is \texttt{diesel\_cli},\footnote{\url{https://crates.io/crates/diesel_cli}} which yanked version \texttt{0.99.0} to restrict the dependency of \texttt{clap} from \texttt{$\ge$2.27.0} to \texttt{$\wedge$2.27.0} to prevent using \texttt{3.x.x} versions of \texttt{clap} since major updates could introduce breaking changes. For \textit{Yanked dependencies}, one example is version \texttt{0.9.2} of \texttt{winping}.\footnote{\url{https://crates.io/crates/winping}} This version was yanked \enquote{due to a yanked dependency}\footnote{\url{https://github.com/TyPR124/winping/blob/master/RELEASES.md}} and \texttt{winping} had to update the dependency of the \texttt{quote}\footnote{\url{https://crates.io/crates/quote}} package from \texttt{1.0.2}~(yanked) to \texttt{1.0.3} in version \texttt{0.9.3}.

\textit{Bump propagation}~(5.1\%) is a notable rationale, which happens when updating the existing dependencies of a package. For instance, \texttt{sdl2}\footnote{\url{https://crates.io/crates/sdl2}} updated the requirement of \texttt{sdl2-sys} from \texttt{$\wedge$0.7.0} to \texttt{$\wedge$0.8.0} and published a patch update \texttt{0.12.2} for \texttt{0.12.1}. Since another package \texttt{sdl2\_image} depends on \texttt{$\wedge$0.12.1} of \texttt{sdl2} and \texttt{$\wedge$0.7.0} of \texttt{sdl2-sys} \enquote{which leads to conflicts},\footnote{\url{https://github.com/Rust-SDL2/rust-sdl2/issues/478}} \texttt{sdl2\_image} was broken and \texttt{sdl2} had to yank \texttt{0.12.2} and republished it as \texttt{0.13.0}. The behaviour of bumping the required version of a dependency~(\texttt{sdl2-sys}) in \texttt{sdl2} broke its dependent (\texttt{sdl2\_image}) and \texttt{sdl2} had to bump its version number as well.

\medskip\noindent\textbf{Pattern 2: Yanking all releases~(25\% of the packages).} We observed this pattern in 1,172 packages, with 60\% of these packages having only one release. The main rationale behind this pattern~(50.0\%) in the card sort is \textit{Package removed or replaced} since developers cannot point a new dependency to a fully yanked package. One example is \texttt{ncollide},\footnote{\url{https://crates.io/crates/ncollide}} the owner explained that \enquote{the overly generic crate ncollide has been replaced by two distinct crates: ncollide2d and ncollide3d which are dedicated to 2D and 3D respectively.}\footnote{\url{https://github.com/dimforge/ncollide/issues/322}} Another example is \texttt{c},\footnote{\url{https://crates.io/crates/c}} which yanked all releases to \enquote{kill}\footnote{\url{https://github.com/hilbert-space/c/issues/1}} this package. We also identified \textit{Breaking SemVer}~(25.0\%) and \textit{Fixing \enquote{*} dependencies}~(25.0\%) rationales under this pattern, however, the identified rationales are not for the whole package, but for some specific releases.

\medskip\noindent\textbf{Pattern 3: Yanking back-to-back releases~(17\% of the packages).} There are 814 packages that followed this pattern and \textit{Defect}~(46.7\%) is the most common rationale that we identified in the card sort under Pattern 3. We noticed that developers often yanked multiple releases due to the same defect. For example, \texttt{clap}\footnote{\url{https://crates.io/crates/clap}} yanked the versions from \texttt{1.4.0} to \texttt{2.21.0} since these versions are all affected by \enquote{an erroneous definition of a macro}.\footnote{\url{https://github.com/clap-rs/clap/issues/2076}} In fact, one of the maintainers left a comment about \texttt{Cargo} not supporting \enquote{yank[ing] everything from X.X.X to Y.Y.Y}. We also observed packages that yanked older releases for security purposes, such as \texttt{untrusted}\footnote{\url{https://github.com/briansmith/untrusted/issues/29}} and \texttt{bitvec}.\footnote{\url{https://github.com/bitvecto-rs/bitvec/issues/59}} Moreover, the only instance of the \textit{Unsupported} rationale in the card sort was found under this pattern. Old releases of \texttt{ring}\footnote{\url{https://crates.io/crates/ring}} were yanked because the owner no longer supports these versions, even though they do not have any known vulnerabilities. Instead, the owner recommends that people only use the latest version.\footnote{\url{https://github.com/briansmith/ring/issues/774}} In addition, we observed that packages yanked multiple releases due to the \textit{MSRV policy}~(6.7\%). For instance, \texttt{block-buffer}\footnote{\url{https://crates.io/crates/block-buffer}} yanked versions \texttt{0.7.0} to \texttt{0.7.2} since these versions used an interface that \enquote{was stabilized only in Rust 1.28},\footnote{\url{https://github.com/RustCrypto/utils/issues/22}} hence, the dependents of \texttt{block-buffer} will be broken if they used a lower version of \texttt{Rust}. 

\medskip\noindent\textbf{Pattern 4: Yanking all releases except one (11\% of the packages).} There are 506 packages that follow this pattern and 94\% of these packages left their newest release unyanked. We only found three packages that explained the rationales under this pattern in the card sort. All three packages explained that at least one of the yanked releases contained a defect. \texttt{battery-cli}\footnote{\url{https://crates.io/crates/battery-cli}} is the only package that explained that the rationale for yanking was \textit{Package removed or replaced}. This package left the newest release unyanked to show the \texttt{readme} file which mentions that \enquote{this crate was yanked and replaced by battop crate} instead of the default page on the website.

\medskip\noindent\textbf{Pattern 5: Yanking nonadjacent releases~(7\% of the packages).} There are 383 packages that do not belong to any patterns which we proposed. Similar to Pattern 1 and Pattern 3, the two most common rationales that we identified in the card sort are \textit{Defect}~(50.0\%) and \textit{Breaking SemVer}~(42.9\%). In addition, the only instance of \textit{License updated} we identified in the sort is from this group. Particularly, \texttt{sic}\footnote{\url{https://crates.io/crates/sic}} yanked version \texttt{0.10.0} since the \enquote{dependency licenses [were] not updated}.\footnote{\url{https://github.com/foresterre/sic/issues/193}} \texttt{sic} also yanked version \texttt{0.7.1} for a defect which \enquote{fails to build from crates.io}.\footnote{\url{https://github.com/foresterre/sic/issues/50}} 


\textbf{5.3\% of the packages with at least one yanked release explain the rationales in their changelogs, issue reports, and pull requests.} During the pattern analysis, we observed that the proportion of packages which explain the rationales behind yanked releases is relatively small, compared to 64\% of the deprecation messages in \texttt{npm} which explain the rationales for deprecating a package or release. Hence, it is not possible to identify the rationales behind all yanked releases. 

\textbf{1.5\% of the packages with at least one yanked release explain the rationale behind yanking a release in a changelog.} There are 21\% (994 packages) of the packages with at least one yanked release that have a changelog, but only 1.5\% (70 packages) of these changelogs explain the rationale behind yanking a release. In other words, if a developer desired to know the rationale behind a yanked release of a package by searching its \texttt{changelog}, only 2\% of the packages can provide an answer for the developer. In addition, we noticed that most (93\%) of the changelogs in the packages with at least one yanked release are kept in a separate file (e.g., \texttt{changelog.md} or \texttt{releases.md}). However, 86\% of the packages which have a \texttt{changelog} file do not mention this file in their \texttt{readme}. As a result, developers could overlook the \texttt{changelog} file if they did not access the \texttt{GitHub} repository of these packages.

\textbf{4.1\% of the packages with at least one yanked release explain the rationale behind yanking a release in the issue reports or pull requests of their GitHub repository.} We found that 192 packages explained the rationale behind yanking a release in their issue reports or pull requests~(13 of the packages also explained in changelogs). In addition, we noticed that the pull requests were usually created by the owner of the package to explain why a release was yanked. However, the issue reports were mostly created by developers who used the package to ask questions about yanked releases.

\hypobox{
	\textbf{RQ2 Summary:} 
	 In \texttt{Cargo}, packages use the yank command for several reasons other than just to indicate a release is defective, but they rarely provide the reason for yanking.
}

%

\subsection{RQ3: How many packages adopt yanked releases?} \label{res:yanked_adoption}

\textbf{Motivation.} \texttt{Cargo} does not allow the owner of a package to delete their releases, but it allows the owner to remove a release from the registry index by yanking. Hence, packages cannot point a new dependency to a yanked release because \texttt{Cargo} cannot find it in the registry index. In this research question, we study how often packages directly adopted yanked releases. 


In addition, we study how many releases have \textit{unresolved dependencies} due to yanked releases. The yank mechanism in \texttt{Cargo} is more forceful than the deprecation mechanism in \texttt{npm}. When \texttt{npm} resolves the dependency requirements for a package, it will use a deprecated release if needed and provide a warning message. However, \texttt{Cargo} will not choose a yanked release even if only this yanked release can satisfy the dependency requirements, which leads to unresolved dependencies. We call a release with unresolved dependencies an \textit{implicitly yanked release}, because \texttt{Cargo} cannot use this release to resolve the dependency requirements of other package (and hence build those packages) even though this release was not yanked explicitly.

\medskip\noindent\textbf{Approach.} We analyzed the \texttt{dependencies} table in the database~(as shown in Table~\ref{tab:database_table}) to collect packages which directly adopted at least one yanked release. First, we selected the dependency requirements which can be satisfied by a yanked release. Then, we collected the information about the owners (i.e., releases from different packages) of these requirements. For each yanked release $d$, we calculated the proportion $p_d$ of direct adoptions of yanked releases~\cite{deprecation_release:Filipe}. The value of $p_d$ was calculated by:

$$ p_d = \frac{a_d}{\sum_{d\in yanked\ releases}{a_d}}$$

\noindent where $a_d$ is the number of times that release $d$ is directly adopted by a package. For example, if packages directly adopted 100,000 yanked releases~(sum of $a_d$) and a yanked release \textit{d} accounts for 1,000 times of these adoptions~($a_d$), the value of $p_d$ is 1\%.

\begin{figure}[t]
	\centering
	\includegraphics[width=2.3in]{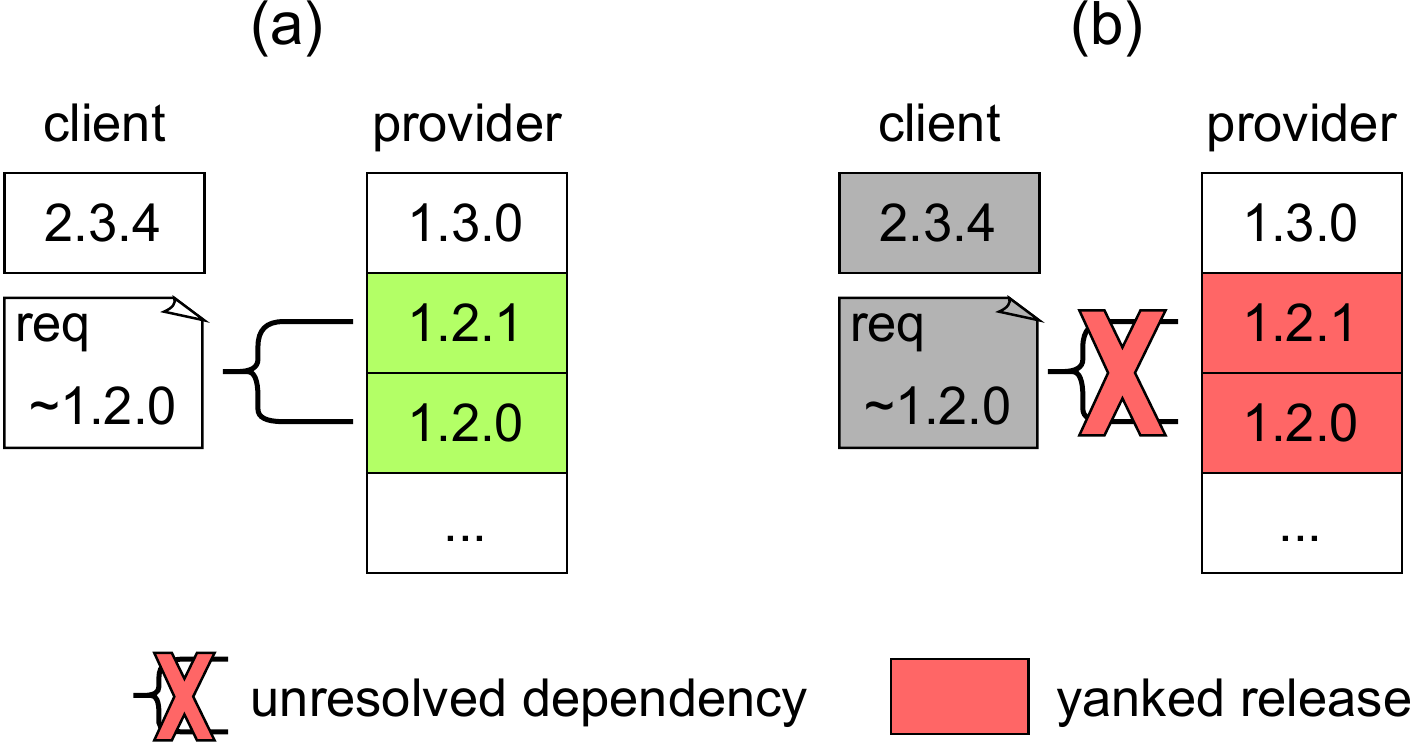}
	\caption{Two scenarios of resolving dependencies: (a)~The dependency requirement can be resolved; (b)~The dependency requirement cannot be satisfied because of yanking.}
	\label{fig:RQ3_approach}
\end{figure}

Next, we study how many releases have unresolved dependencies because of adopting yanked releases. Figure~\ref{fig:RQ3_approach} shows two scenarios in which a dependency requirement of a client package points to a provider package. In the first scenario, version \texttt{2.3.4} of the client package has a requirement \texttt{$\sim$1.2.0}, and this requirement can be resolved by \texttt{1.2.0} or \texttt{1.2.1} of the provider package. In the second scenario, only versions \texttt{1.2.0} and \texttt{1.2.1} can satisfy the requirement \texttt{$\sim$1.2.0} but these two versions are both yanked. We investigated all the dependency requirements and collected the releases which have unresolved dependencies. 


In addition, we study how yanked releases propagate through the dependencies in the ecosystem. The propagation happens when yanked releases break dependency requirements and cause implicitly yanked releases~(i.e., releases with unresolved dependencies). We collected the implicitly yanked releases which directly adopted yanked releases. However, the propagation continues if those implicitly yanked releases cause new unresolved dependencies. Hence, we performed the analysis recursively to collect all implicitly yanked releases. 

\begin{figure}[t]
	\includegraphics[width=3.2in]{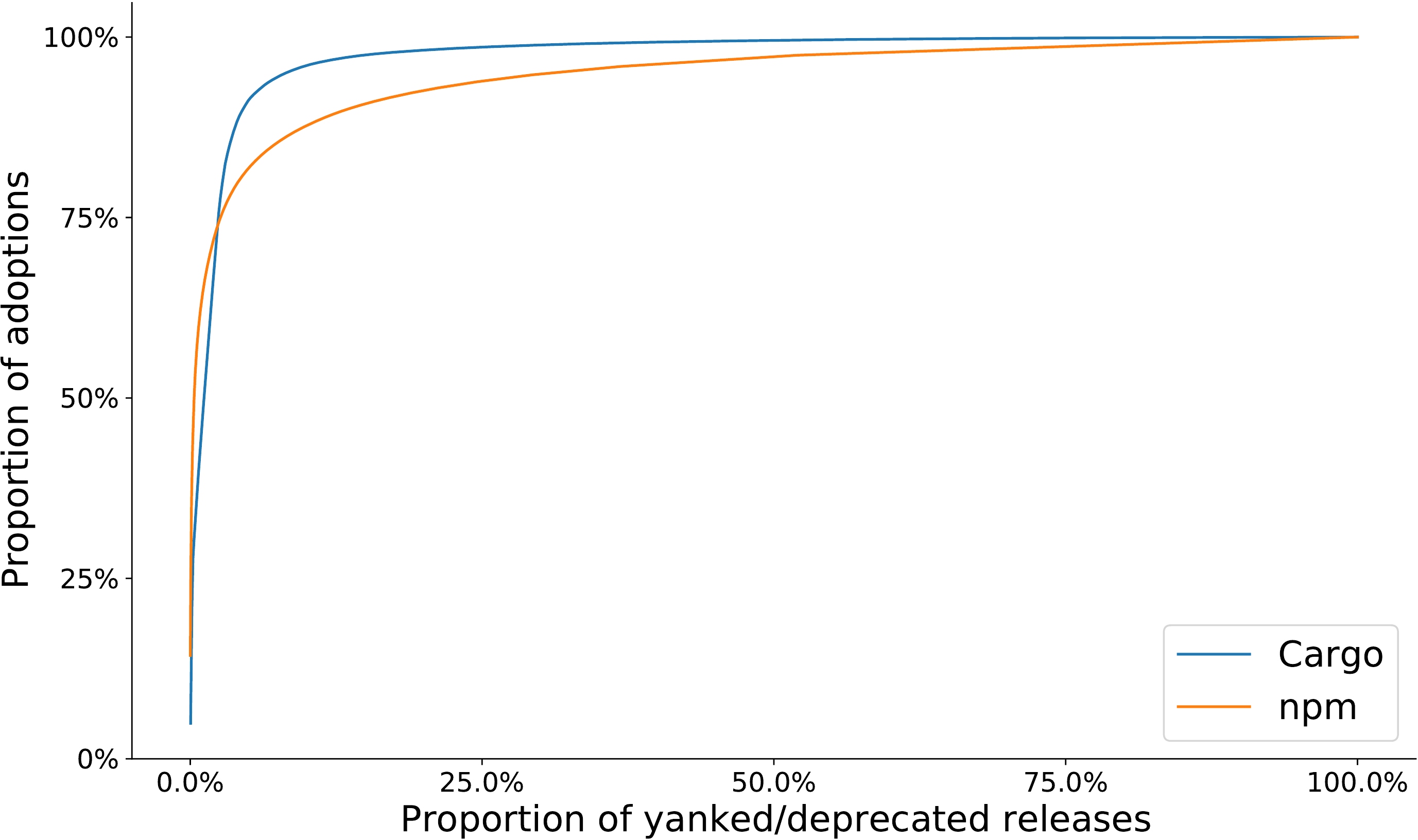}
	\caption{Cumulative histogram for the proportion of direct adoptions of yanked releases in \texttt{Cargo} and \texttt{npm}.}
	\label{fig:yanked_vs_adopts}
\end{figure}

\medskip\noindent\textbf{Findings.} \textbf{46\% of packages in Cargo directly adopted at least one yanked release and 2.4\% of the yanked releases accounted for 75\% of these adoptions.} There are 22,277 packages that directly adopted at least one yanked release of a partially yanked package and 268 packages directly adopted fully yanked packages in \texttt{Cargo}. The proportion of packages which directly adopted yanked releases in \texttt{Cargo} is 19\% higher than in \texttt{npm} (27\%). Although such adoptions do not directly put a package at risk (since there could be newer releases that satisfy a requirement), they could contribute to unresolved dependencies if another release is yanked. One reason could be that \texttt{Cargo} has a higher proportion of packages that are partially yanked~(Section~\ref{res:how_many_yanked}). In addition, Figure~\ref{fig:yanked_vs_adopts} shows that most of the direct adoptions of yanked releases are concentrated on a relatively small proportion of yanked releases in both \texttt{Cargo} and \texttt{npm}.

\textbf{1.4\% of the releases in \texttt{Cargo} have unresolved dependency requirements which are related to yanked releases.} 4,158 releases in \texttt{Cargo} have unresolved dependency requirements because it directly or transitively adopted a yanked release and these releases became implicitly yanked releases. We found that 65.2\%~(2,712 releases) of the implicitly yanked releases are caused by packages that follow Pattern 3~(yanking back-to-back releases) in Section~\ref{res:yanked_in_practice} and 39.2\%~(1,631 releases) of the implicitly yanked releases are caused by the \texttt{ring} package which yanked unsupported old releases. Moreover, we noticed that 15 releases from 5 packages have invalid dependency requirements~(i.e., they are unrelated to yanked releases). For example, 10 releases of \texttt{leveldb},\footnote{\url{https://crates.io/crates/leveldb}} version numbers \texttt{0.3.4} to \texttt{0.5.1}, have a common dependency requirement that indicates that a version of \texttt{db-key}\footnote{\url{https://crates.io/crates/db-key}} should satisfy the constraint \texttt{$\wedge$0.0.4}. However, there is no release (i.e., yanked or unyanked) in \texttt{db-key} that can fulfill this requirement.



\textbf{54\% of the implicitly yanked releases are caused by direct adoption of yanked releases.} There are 2,266 releases that directly adopted yanked releases and became implicitly yanked releases. For example, version \texttt{0.1.2} of \texttt{chrono}\footnote{\url{https://crates.io/crates/chrono}} has a caret dependency constraint for \texttt{time}\footnote{\url{https://crates.io/crates/time}} (i.e., \texttt{$\wedge$0.0.3}), but \texttt{time} yanked all releases before \texttt{0.1.0}. In this case, this dependency constraint cannot be satisfied and version \texttt{0.1.2} of \texttt{chrono} becomes an implicitly yanked release. We found that 54\% (1,411 releases) of these implicitly yanked releases were caused by only 10 packages. For example, \texttt{clap},\footnote{\url{https://crates.io/crates/clap}} is a library for parsing command-line arguments which yanked 91 (out of 211) releases. There are 353 releases from 85 packages which directly adopted at least one of these yanked releases from \texttt{clap} and became implicitly yanked.



\begin{figure}[t]
	\includegraphics[width=3.4in]{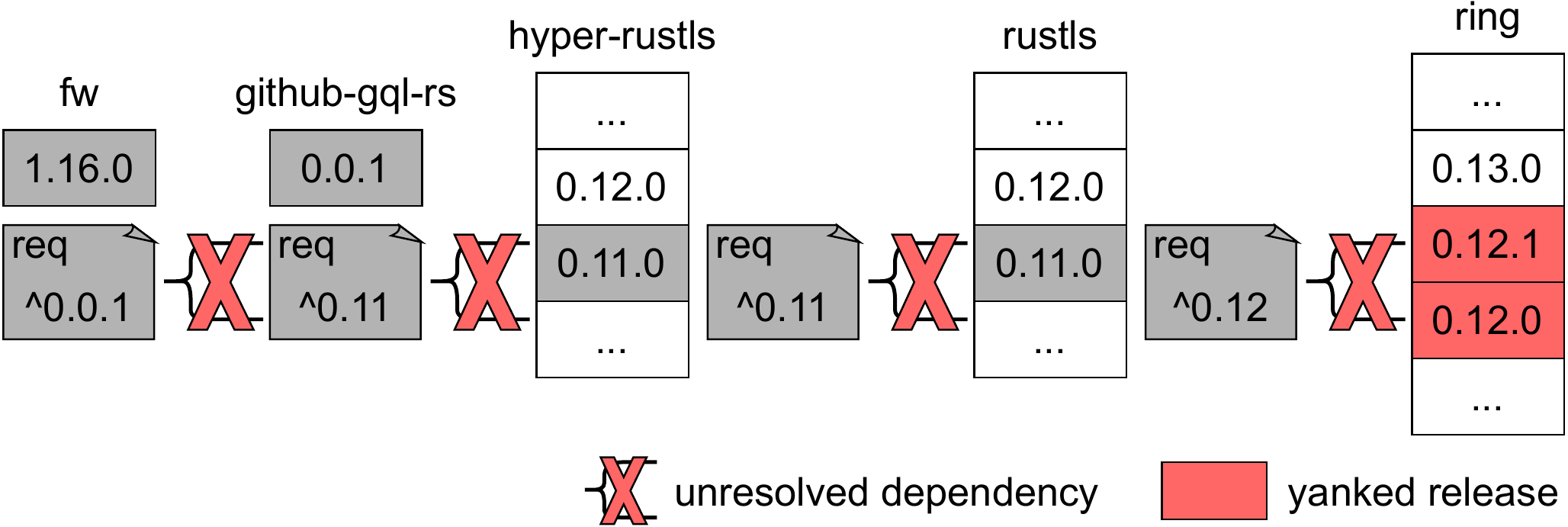}
	\caption{An example of yanking propagation of \texttt{ring}. \texttt{0.11.0} of \texttt{rustls}, \texttt{0.11.0} of \texttt{hyper-rustls}, \texttt{0.0.1} of \texttt{github-gql-rs}, and \texttt{1.16.0} of \texttt{fw} became implicitly yanked releases.}
	\label{fig:yanked_propagated}
\end{figure}

\textbf{46\% of the implicitly yanked releases are caused by the transitive adoption of a yanked release and 54\% of them are propagated from a single package.} We found 1,892 implicitly yanked releases that transitively adopted yanked releases, and then themselves became implicitly yanked. Figure~\ref{fig:yanked_propagated} shows that version \texttt{1.16.0} of \texttt{fw}\footnote{\url{https://crates.io/crates/fw}} became implicitly yanked since it transitively adopted yanked releases of \texttt{ring}.\footnote{\url{https://crates.io/crates/ring}} In this case,
the implicitly yanked release \texttt{0.11.0} of \texttt{rustls}\footnote{\url{https://crates.io/crates/rustls}} was propagated from two yanked releases of \texttt{ring}. Then, the propagation happened to \texttt{hyper-rustls},\footnote{\url{https://crates.io/crates/hyper-rustls}} \texttt{github-gql-rs},\footnote{\url{https://crates.io/crates/github-gql-rs}} and \texttt{fw}. The depth of this propagation is 4 and we noticed that the maximum depth of yanking propagation in \texttt{Cargo} is also 4. However, 1,631~(39\%) out of 4,158 implicitly yanked releases were propagated from a common root package \texttt{ring}.

\hypobox{
	\textbf{RQ3 Summary:} 
	46\% of packages directly adopted at least one yanked release. Yanked releases were propagated in \texttt{Cargo} and 1.4\% of the releases that are currently in \texttt{Cargo} have unresolved dependencies due to the yanking propagation.
}


\section{Implications} \label{sec:implications}

In this section, we discuss our findings and implications for the maintainers of package managers, the package owners, the maintainers of \texttt{Cargo}, and researchers.


\subsection{Implications for maintainers of package managers}

\textbf{Package managers should implement a release-level deprecation mechanism.}  The proportion of deprecated releases in \texttt{Cargo} and \texttt{npm} are 3.7\% and 3.2\% respectively, and the percentage of deprecated releases has gradually increased from 2014 to 2020 in Cargo  (Section~\ref{res:how_many_yanked}). Since the release-level deprecation mechanism has seen increased use in these two ecosystems, there is likely a need for it in other ecosystems as well. In recent years, \texttt{PyPI} and \texttt{NuGet} saw the need and have started to support release-level deprecation in April 2020 and September 2019. We suggest other package managers also implement the release-level deprecation mechanism. 

\medskip\noindent\textbf{Package managers should provide features to support the various ways in which developers deprecate releases.} We observed five patterns of yanking in \texttt{Cargo}~(Section~\ref{res:yanked_in_practice}). To support Pattern 2, we suggest that the deprecation mechanism of package managers should support package-level deprecation, which is also used by 80\% of the deprecations in \texttt{npm}~\cite{deprecation_release:Filipe}. To support other patterns, we suggest package managers to support deprecating a single release or multiple releases at the same time to offer flexibility for developers. 

\medskip\noindent\textbf{Package managers should allow package owners to decide whether a deprecation is forceful or not.} A forceful deprecation mechanism like yanking in \texttt{Cargo} can cause unresolved dependencies for releases of other packages~(Section \ref{res:yanked_adoption}). We suggest that package managers leave the choice of forceful deprecation to the package owners. The package owners can deprecate a release forcefully~(like in \texttt{Cargo}) when it is necessary, such as if they found security vulnerabilities in a cryptography package. Otherwise, the package owner can decide to deprecate a release non-forcefully~(like in \texttt{npm}) and allow developers to decide whether they still want to adopt the deprecated release in their own packages. In addition, we suggest that package managers provide a warning with information about how many packages would break~(similar to our analysis in Section \ref{res:yanked_adoption}) if a developer decides to deprecate forcefully.

\subsection{Implications for package owners}
\textbf{Package owners should explain the rationales behind yanked releases in the documentation.} 94.7\% of the packages that have at least one yanked release in Cargo never explained the rationale behind yanking~(Section~\ref{res:yanked_in_practice}). This percentage is high compared to \texttt{npm} in which for 64\% of the deprecated releases a rationale is known~\cite{deprecation_release:Filipe}. Since \texttt{Cargo} does not support adding a message for a yanked release, we recommend that package owners record the reason for yanking a release in the package's documentation. For example, package owners can create an issue report to track a yanked release and put its link into the readme or changelog. The issue report should contain detailed information about a yanked release, and provide a place for developers to discuss this release, or to give advice on how to deal with the yanking. In addition, we noticed that only 21\% of the packages (Section~\ref{res:yanked_in_practice}) have a changelog in \texttt{Cargo}. We suggest package owners maintain a changelog to tell developers about the notable changes in each release, which can also be used to explain the rationale behind yanked releases.


\medskip\noindent\textbf{Package owners should avoid yanking old releases which are no longer supported without providing an alternative release or migration guidelines.} We found that 39.2\% of the implicitly yanked releases in \texttt{Cargo} are caused by the \texttt{ring} package which yanked old unsupported releases~(Pattern 3 in Section~\ref{res:yanked_in_practice}), even though these packages had no known vulnerabilities~(Section~\ref{res:yanked_adoption}). For the packages that insist on yanking unsupported releases, we recommend they indicate replacement releases or provide guidelines for developers to migrate away from yanked releases. 


\subsection{Implications for Cargo maintainers}
We compare the yanked mechanism in \texttt{Cargo} with the deprecation mechanism in \texttt{npm} based on our findings. We summarize the difference in Table~\ref{tab:comparision_npm_cargo} for \texttt{Cargo} maintainers.


\begin{table}[t]
	\caption{Comparisons of the yanked mechanism in Cargo and the deprecation mechanism in npm.}
	\label{tab:comparision_npm_cargo}
	\begin{tabularx}{\linewidth}{XXXXX}
		\toprule
		\textbf{Package manager} & \textbf{Record deprecation date} & \textbf{Describe rationale} & \textbf{Deprecate a package} & \textbf{Ban yanked releases} \\
		\midrule
		\texttt{Cargo} & \checkmark & & & \checkmark \\
		\texttt{npm}   & & \checkmark  & \checkmark   & \\
		\bottomrule
	\end{tabularx}
\end{table}

\medskip\noindent\textbf{Cargo should allow the owner of a package to add a note to a yanked release and provide a warning for packages that adopted it.} We found that the percentage of packages that explain the rationales behind yanked releases is low in \texttt{Cargo}~(Section~\ref{res:how_many_yanked}) compared to \texttt{npm}~(5.3\% vs. 64\%). One reason could be that \texttt{npm} allows the package owners to add a message for deprecated releases while \texttt{Cargo} does not. Moreover, we observed that it is much more common in \texttt{Cargo} to yank only a few releases instead of the whole package~(Section~\ref{res:yanked_in_practice}) and the owners of packages yanked releases for various reasons. For fully yanked packages, at least developers know that these packages will probably no longer be maintained. However, developers who depend on a partially yanked package can hardly understand what is happening since there is no mechanism for describing why a release was yanked. Hence, we recommend that \texttt{Cargo} should allow the owner of a package to leave a message when they are yanking a release. There is an issue report\footnote{\url{https://github.com/rust-lang/cargo/issues/2608}} asking for the same functionality since April 23, 2016, but it is still not implemented. With the increasing proportion of yanked releases in the ecosystem (Section~\ref{res:how_many_yanked}), more developers will be affected by this issue. In addition, we recommend that \texttt{Cargo} should provide a warning message for packages that adopted a yanked release.


\medskip\noindent\textbf{Cargo should detect implicitly yanked releases and provide a warning for these releases.} We found that 1.4\% of releases in \texttt{Cargo} are implicitly yanked~(Section~\ref{res:yanked_adoption}). We recommend that \texttt{Cargo} should mention that a release is implicitly yanked on the webpage of a package. For example, there could be an \enquote{unresolved} label beside an implicitly yanked release on the webpage, hence developers can avoid using this release. In addition, for a package which adopted an implicitly yanked release, \texttt{Cargo} can show the dependency tree and indicate the release which breaks the dependencies for developers.

\medskip\noindent\textbf{Cargo should warn the owner of a package which adopts a yanked release in its lock file.} \texttt{Cargo.lock} stores the information about dependencies locally for a project if the project was compiled successfully. However, one of the dependencies can be yanked after the compilation and \texttt{Cargo} does not inform the developer. Since the proportion of packages which directly adopted yanked releases is 46\% in \texttt{Cargo}~(Section~\ref{res:yanked_adoption}), we recommend that \texttt{Cargo} should check up the package registry when developers are building their project based on the \texttt{Cargo.lock}. Hence, \texttt{Cargo} can give a warning message to developers if one of the dependencies was yanked.


\subsection{Implications for researchers}
\textbf{Researchers should study automatic semantic versioning guarantee checkers to detect whether a release follows the guarantee.} In Section~\ref{res:yanked_in_practice}, we found that \textit{Breaking SemVer} is the most common rationale behind yanking. This finding indicates the difficulty for package owners to decide whether an update follows the guarantee. Automatic semantic versioning guarantee checkers can help package owners by analyzing the code before a release is published. In addition, these checkers can be used to analyze how packages in different software ecosystems follow the semantic versioning guarantee.



\section{Threats to validity} \label{sec:validity}

In this section, we discuss the threats to the validity of our study about yanked releases in \texttt{Cargo}.

\textit{Internal validity:}
We analyzed the percentage of yanked releases from 2014 to 2020 based on the \texttt{Git} history of the registry index. Ideally, this index repository is updated automatically by a program. However, we found eight records which show that the maintainer edited the index manually to delete some packages. Since these deletions are not considered, our results will include the yanked releases of these deleted packages.

The identification of changelogs is based on searching keywords in \texttt{readme} and matching filenames under the root directory of packages' \texttt{GitHub} repository. It is possible that the owner of a package did not use a word in our keyword list to indicate their changelog. We randomly selected 100 readmes from our dataset and manually identified 7 readmes that contain a changelog. The result is exactly the same as our heuristic approach, hence, this heuristic approach is considered reliable for our dataset. 

In addition, we filter out changelogs, issue reports, and pull requests which do not have \enquote{yank} and \enquote{deprecate} in the content. This filtering might exclude information that explains why a release was yanked. We randomly selected 100 changelogs and 100 issue reports/pull requests from the excluded samples to manually verify whether they contain the rationales for yanked releases. We found that the keywords searching approach missed 3 of the 100 changelogs~(2 for \textit{Package removed or replaced} and 1 for \textit{Defect}) and 0 of the 100 issue reports/pull requests. Since few packages maintained a changelog and most rationales are identified from issue reports/pull requests (Section~\ref{res:yanked_in_practice}), the result of identifying rationales is considered reliable. Besides, we selected 638 packages which have the keywords in changelogs, issue reports, or pull requests and filtered out 380 out of the 638 for card sorting. The author who did not participate in the card sorting independently analyzed 100 samples of the 380 packages and reported the false negative rate is 9\%.

\textit{External validity:}
We studied yanked releases in the \texttt{Rust} and \texttt{npm} package registries, but the findings of our study may not generalize to the package managers of other programming languages. One reason could be that other package managers may not provide a release-level deprecation mechanism for developers, and the identification of whether a release is yanked could be complicated (or not possible). Future studies should further investigate other software ecosystems with release-level deprecation.

We investigated the changelogs, issue reports, and pull requests of packages with at least one yanked release. However, we only looked at packages which provide a link to their \texttt{GitHub} repository. There are 15\% of these packages that do not have a link to their repository and 214 packages (4\%) provide a link to other repository hosting platforms such as \texttt{Bitbucket}\footnote{\url{https://bitbucket.org/}} and \texttt{GitLab}.\footnote{\url{http://gitlab.com/}} Hence, future studies are necessary to investigate if our results hold for packages that maintain their code outside of \texttt{GitHub}.

The results of our study might not apply directly to other software ecosystems, because the community and the development model of the programming language can also affect the results. However, our methodology can be applied to analyze other software ecosystems.

\section{Conclusion} \label{sec:conclusion}


We studied 48,823 packages in \texttt{Cargo} to understand the yank mechanism. In particular, we studied the frequency in which the yanked mechanism is used, the patterns of yanking in packages and the rationales behind yanking, and the adoption of yanked releases. The most important findings of our study are:

\smallskip\begin{enumerate}[1.]
	\item The proportion of yanked releases in \texttt{Cargo} has increased from 1.4\% to 3.7\% between 2014 and 2020.
	\item Even though the proportions of packages with at least one yanked release are similar in \texttt{Cargo} and \texttt{npm}, these packages in \texttt{Cargo} have a lower yanking rate.
	\item We observed 5 yanking patterns in \texttt{Cargo} and the rationales include withdrawing a defective release or a release that broke the semantic versioning guarantee, indicating a package is removed or replaced, or fixing dependencies.
	\item 46\% of packages directly adopt at least one yanked release in \texttt{Cargo}. These yanked releases were propagated through the adoption, causing 1.4\% of the releases in \texttt{Cargo} to have unresolved dependencies and hence become unbuildable.
\end{enumerate}

In this paper, we found that yanked releases cause unresolved dependencies since the yank mechanism is more forceful in \texttt{Cargo} than \texttt{npm}. Based on our findings, we suggest that \texttt{Cargo} should provide a package-level deprecation mechanism and allow package owners to leave a reason for yanking a release, and we recommend that other package managers integrate a release-level deprecation mechanism as well. 

\ifCLASSOPTIONcompsoc
  \section*{Acknowledgments}
\else
  \section*{Acknowledgment}
\fi
The work described in this paper has been supported by the ECE-Huawei Research Initiative (HERI) at the University of Alberta.




\bibliographystyle{IEEEtranS}
\bibliography{IEEEabrv,mybib}

%



\clearpage

\section*{List of URLs}
\printendnotes[custom]


\end{document}